\documentclass[11pt,letterpaper]{article}
\usepackage[utf8]{inputenc}
\usepackage[english]{babel}
\usepackage{amsmath}
\usepackage{amsfonts}
\usepackage{amssymb}
\usepackage{graphicx}
\usepackage[left=3cm,right=3cm,top=2cm,bottom=3cm]{geometry}
\author{Jorge Pinochet}
\title{\textbf{The Hawking temperature, the uncertainty principle and 
quantum black holes}}
\begin{document}

\author{Jorge Pinochet\\ \\
 \small{\textit{Facultad de Educación}}\\
 \small{\textit{Universidad Alberto Hurtado, Erasmo Escala 1825, Santiago, Chile.} japinochet@gmail.com}\\ \\}

\date{\small \today}
\maketitle

\begin{abstract}
\noindent In 1974, Stephen Hawking theoretically discovered that black holes emit thermal radiation and have a characteristic temperature, known as the \textit{Hawking temperature}. The aim of this paper is to present a simple heuristic derivation of the Hawking temperature, based on the Heisenberg uncertainty principle. The result obtained coincides exactly with Hawking's original finding. In parallel, this work seeks to clarify the physical meaning of Hawking's discovery. This article may be useful as pedagogical material in a high school physics course or in an introductory undergraduate physics course. \\ \\


\end{abstract}

\maketitle

\section{Introduction}

Black holes are one of the most enigmatic and surprising predictions of general relativity, the theory of gravity proposed by Albert Einstein in 1915 [1, 2]. According to general relativity, a black hole is a region of space in which the gravitational field is so intense that nothing can escape from its interior, not even light [3]. A black hole absorbs everything but emits nothing. However, as the British physicist Stephen Hawking discovered, quantum mechanics dramatically changes this paradigm. In a short article published in 1974 and extended in 1975, Hawking theoretically proved that black holes emit thermal radiation and have an absolute temperature, known as the \textit{Hawking temperature}, $T_{H}$ [4, 5]. According to Hawking's calculations, thermal radiation is only significant for so-called \textit{quantum black holes}, which size is smaller than that of an atomic nucleus.\\

Although it has been more than four decades since the British physicist's discovery, his seminal work remains valid, and has led to a broad field of research that is at the forefront of contemporary physics. To demonstrate that black holes have a temperature and emit radiation, Hawking employed very sophisticated tools, and a detailed technical presentation of these findings is therefore only accessible to a small circle of specialists. The aim of this paper is to offer a simple, rigorous and updated exposition of Hawking's findings, aimed mainly at a non-specialist public with a general background in physics and mathematics. The central element of this work is a simple heuristic derivation of $T_{H}$, based on a physical image that is different from that popularised by Hawking in his popular books. The Hawking image uses a phenomenon called quantum vacuum fluctuation, which description is complex, partly because it assumes that the radiation does not come directly from within the black hole, but outside it\footnote{Some specialists have suggested that Hawking's image is not entirely correct. The interested reader can find a technical discussion in [6]; an updated analysis of the topic is found in [7].}. Conversely, the image developed in this work, which is based on the Heisenberg uncertainty principle, assumes that the Hawking radiation comes from the black hole itself. This provides a simple and direct mathematical description of $T_{H}$ and the associated thermal radiation. One of the most remarkable aspects of this new image is that it leads to a value of $T_{H}$ that exactly matches Hawking's original result. However, this new image is not intended to be an accurate description of the mechanism of emission of Hawking radiation; it only aims to be a pedagogical resource that facilitates the understanding of the subject. This work presupposes only a basic knowledge of mathematics and physics. From this perspective, the article may be useful as pedagogical material in a high school physics course or in an introductory undergraduate physics course.\\

In Section 2, an introduction to the concepts of black hole and Hawking temperature is presented. Section 3 introduces the heuristic derivation of Hawking temperature, and discusses the possibility of detecting it. Finally, in Section 4, the meaning, scope and limits of Hawking's finding are analysed within the framework of contemporary theories that seek to unify general relativity and quantum mechanics.

\section{Classical black holes versus quantum black holes}

The physics of the last century was dominated by two great theories: quantum mechanics and the theory of general relativity. The former describes the non-gravitational interactions between small, light objects, such as atoms or elementary particles, where general relativity is negligible, while the latter describes the gravitational interactions between large, heavy objects, such as planets, stars or galaxies, where quantum mechanics is negligible. The classical concept of a black hole was born within the framework of general relativity. Unlike quantum holes, \textit{classic black holes} are large, heavy objects for which quantum mechanics and Hawking's findings have virtually no importance.\\

According to general relativity, a classic black hole is formed when a high concentration of mass, or its equivalent in energy, occurs within a closed spherical region of space called the \textit{horizon}. The gravitational field within the horizon is so intense that light cannot escape from within and is trapped forever. Since according to relativistic physics, nothing in the universe can move faster than light, then no form of matter or energy enclosed within the horizon can cross to the outside. The radius characterising the horizon of a spherically symmetric black hole without rotation and electrically neutral, called \textit{static black hole}, can be calculated using an equation found in 1916 by the German astronomer Karl Schwarzschild [3, 8]:

\begin{equation}
R_{S}=\dfrac{2GM_{BH}}{c^{2}} \sim \left(10^{-27}m\cdot kg^{-1}\right)M_{BH}  
\end{equation}

\begin{figure}
  \centering
    \includegraphics[width=1\textwidth]{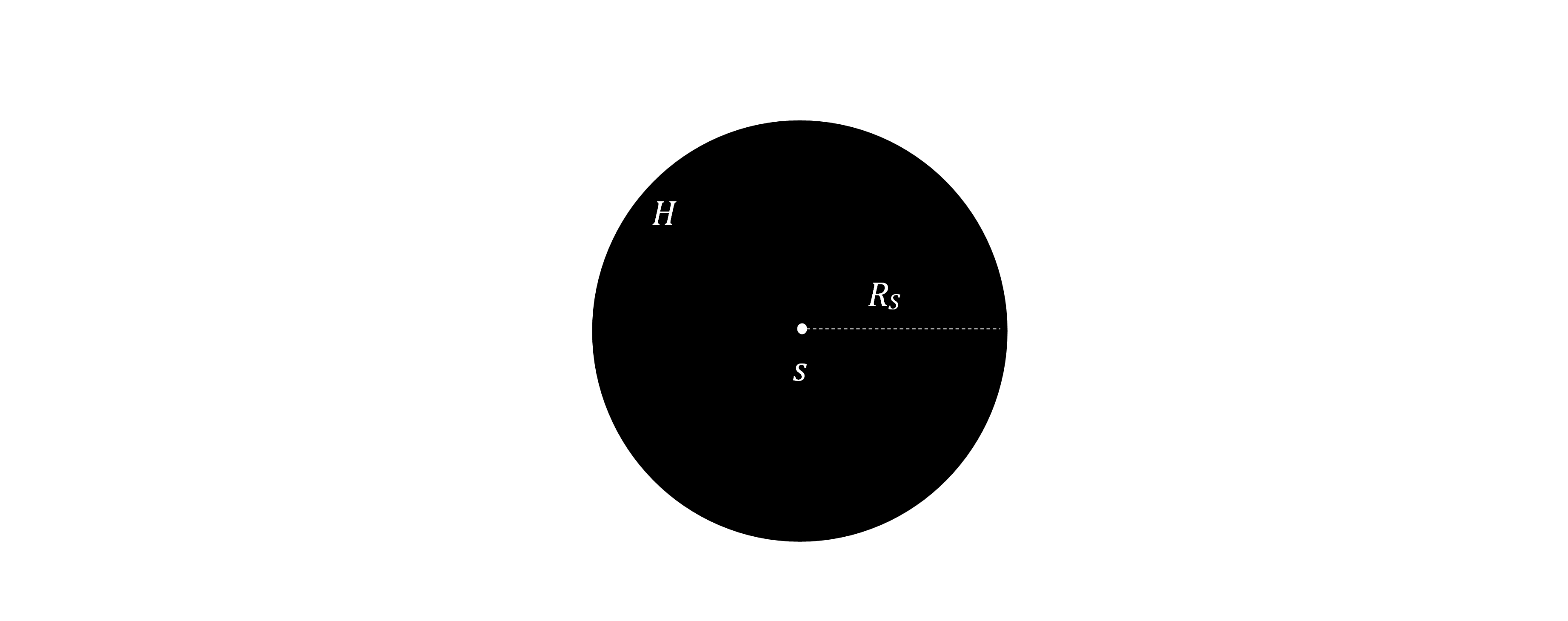}
  \caption{A static black hole. The horizon (\textit{H}) is at a distance $R_{S}$ from the singularity (\textit{S}).}
\end{figure}

where $R_{S}$ is known as \textit{Schwarzschild radius}\footnote{There is an intuitive Newtonian argument for obtaining Equation (1). The escape velocity from the surface of a massive object of radius $R$ and mass $M$ is $V_{e} = (2GM/R)^{1/2}$. Taking $V_{e} = c$ and solving for $R$, we get $R = 2GM/c^{2}$. The physical meaning of this expression is clear: no form of matter or energy contained within the closed spherical surface limited by $R$ can escape, since it would need a speed greater than $c$.}, $G = 6.67\times 10^{-11}N\cdot m^{2}\cdot kg^{-2}$ is the universal gravitation constant, $c=3\times10^{8} m\cdot s^{-1}$ is the speed of light in vacuum, and $M_{BH}$ is the mass of the black hole. To get an idea of the colossal concentrations of matter involved in the formation of a black hole, note that if the earth’s mass, $\sim10^{24}kg$, is put into Equation (1), we obtain $R_{S} \sim 10^{-3}m$, which is less than the size of a marble.\\

General relativity ensures that once the mass-energy has contracted to within the horizon, this contraction continues with nothing to stop it until everything is reduced to a mathematical point of infinite density called the \textit{singularity}, located at the centre of the horizon (see Figure 1). Although the horizon is not a physical surface, it can be visualised as a unidirectional membrane that only allows matter or energy to flow inward [9]. In particular, the horizon must have zero temperature, since according to the laws of thermodynamics, any object with a temperature above absolute zero must emit radiation.\\

Suppose we reduce the mass of a black hole enough so that $R_{S} \sim 10^{-15} m$, which corresponds to the radius of an atomic nucleus. It is evident that we are entering the domain of quantum mechanics, since this is an extremely small size. What is the mass of this black hole? A simple calculation using Equation (1) tells us that if $R_{S} \sim 10^{-15} m$, then $M_{BH} \sim 10^{12} kg$. This figure is a factor $10^{39}$ greater than the mass of a typical atomic nucleus, and is roughly equivalent to the mass of an asteroid or a mountain. This is a heavy object, and we are therefore now in the domain of general relativity. Which theory should we then use to describe a black hole with $R_{S} \sim 10^{-15} m$ and $M_{BH} \sim 10^{12} kg$? General relativity or quantum mechanics? \\

The answer was given by Hawking in his seminal articles of 1974 and 1975: we must use both theories, since this is a small and heavy object, that is, a quantum black hole [4, 5]. In more precise terms, Hawking showed that by combining general relativity, quantum mechanics and thermodynamics\footnote{Strictly speaking, thermodynamics is included in quantum mechanics through so-called quantum statistics. However, for pedagogical purposes, it is clearer to separate quantum mechanics from thermodynamics.}, it follows that a static black hole located in a vacuum\footnote{The fact that the hole is in a vacuum is an important aspect of Hawking's finding; this means that the emission of radiation does not depend on mechanisms related to the presence of material outside the horizon, as is the case with accretion, a physical process that generates large radiation emissions in black holes that are part of binary systems.}  must emit from its horizon in all directions a type of thermal radiation known as \textit{Hawking radiation}. According to Hawking's calculations, for an observer located at a large distance from the horizon (ideally infinite), this radiation has a blackbody spectrum whose absolute temperature is Hawking temperature:

\begin{equation}
T_{H} = \dfrac{\hbar c^{3}}{8\pi kGM_{BH}}
\end{equation}

where $\hbar =h/2\pi =1.05\times10^{-34} J\cdot s$ is the reduced Planck constant, and $k = 1.38\times 10^{-23} J\cdot K^{-1}$ is the Boltzmann constant. As with any hot body, a black hole emits mainly photons. However, if $T_{H}$ is high enough, it is possible that other particles such as neutrinos, electrons, protons etc. are emitted, and a black hole therefore radiates at the cost of reducing its own mass-energy. Equation (2) reveals that as $M_{BH}$ decreases, $T_{H}$ increases, and the radiation becomes more intense. As a result, the mass decreases more and more rapidly, in a process called \textit{evaporation}.\\ 

If we follow evaporation to its ultimate consequences, we see that according to Equation (2), $T_{H}$ should become infinite in the last moments of life of the black hole, when $M_{BH}$ tends to zero. This result suggests that Equation (2) has no general validity. This problem is one of the great challenges of contemporary theoretical physics, and a satisfactory solution still has not been found [10]. However, for reasons of consistency with other physical theories that have been widely confirmed, specialists agree that Equation (2) is correct, provided that $M_{BH}$ is not too small.\\

Following the procedure employed by Hawking in his articles of 1974 and 1975, but at an elementary level, the next section presents a heuristic derivation of Equation (2) that combines basic notions of gravitation, quantum theory and thermodynamics.

\section{Hawking temperature and the uncertainty principle} 
In the previous section, it was pointed out that in order to cross the horizon towards the exterior, a speed greater than that of light in a vacuum, $c$, is required, which is forbidden by relativistic physics. However, the Heisenberg uncertainty principle allows a particle to move with superluminal rapidity and escape from the quantum hole. How is this possible? Before giving a detailed quantitative response, we try to address this question intuitively. Consider a particle of mass $m$. According to the Heisenberg uncertainty principle, if $\Delta x$ is the uncertainty in the position of the particle, and if $\Delta v$ is the uncertainty in its speed, the minimum value that the product of these quantities can take is [11, 12]: 

\begin{equation}
\Delta x \Delta v = \dfrac{\hbar}{2m}
\end{equation}

According to relativistic physics, the mass $m$ and energy $E$ of a particle are related as $E=mc^{2}$, so that:

\begin{equation}
\Delta x \Delta v = \dfrac{c^{2} \hbar}{2E}
\end{equation}

Equations (3) and (4) show that $\Delta x$ and $\Delta v$ are inversely proportional. Since a quantum hole is small, then if a particle is within its interior, the uncertainty in its location, $\Delta x$, is also small. This implies that the uncertainty in the speed can be made large enough so that $\Delta v > c$. When this happens, the particle can escape from the hole and cross the horizon to the outside. Since quantum phenomena are random, it is not possible to determine with certainty which type of particle will escape or at what instant it will do so. The net effect over time of this random process is a continuous flow of particles, which are emitted in all directions from the horizon. This is the \textit{Hawking radiation}\footnote{It is interesting to note that this image of Hawking radiation can be interpreted as being due to the so-called tunnel effect, which is one of the most surprising and counterintuitive consequences of quantum mechanics. The interested reader can find a technical discussion in [13].}.\\

Each particle lost to Hawking radiation removes mass-energy from of the black hole, leading to a reduction in $M_{BH}$ and the evaporation process mentioned in the previous section. Since the emission of this radiation occurs within the range of uncertainty in the speed, it is impossible to observe a particle moving faster than light. Thus, in practical terms the described mechanism does not entail a conflict with the laws of relativistic physics.\\

The above description can be developed in quantitative terms. Imagine a particle inside a static black hole. We know that the uncertainty $\Delta x$ in the location of the particle is determined by the size of the black hole. Since there is a margin of freedom to choose the magnitude of this uncertainty, we can assume that $\Delta x$ is the size of the circumference of the horizon:

\begin{equation}
\Delta x \approx 2\pi R_{S} = \dfrac{4\pi GM_{BH}}{c^{2}}
\end{equation}

The critical condition for the particle to escape from the black hole is $\Delta v = c$. Imposing this condition on Equation (4) and introducing the result into Equation (5):

\begin{equation}
E = \dfrac{c^{3} \hbar}{8\pi GM_{BH}}
\end{equation}

We know that Hawking radiation has a thermal spectrum. According to statistical thermodynamics, this condition can be expressed as: 

\begin{equation}
E \approx kT
\end{equation}

where $T$ is the absolute temperature. Removing $E$ from Equations (6) and (7): 

\begin{equation}
T = T_{H} = \dfrac{\hbar c^{3}}{8\pi kGM_{BH}} \sim \left(10^{23} K \cdot kg\right) M_{BH}^{-1}
\end{equation}

This relation is identical to Equation (2) for the Hawking temperature. However, among others, the assumption $\Delta x \approx 2\pi R_{S}$ introduced in the Equation (5) is not determined solely by physical arguments, since we could also have assumed, for example, that $\Delta x \approx R_{S}$. In any case, the coincidence between Equations (2) and (8) is remarkable, especially considering the simplicity of the derivation.\\

Although we have calculated $T_{H}$ for the case of a quantum black hole, Equation (8) has general validity, since all black holes have temperature and emit thermal radiation, regardless of their mass. However, as Equation (8) reveals, $T_{H}$ only has a significant value when a black hole is small and not very massive. We illustrate this idea through some calculations. If in Equation (8) we use the characteristic mass of a quantum hole, $M_{BH} \sim 10^{12} kg$, we find that $T_{H} \sim 10^{11} K$, a very high temperature that exceeds by a factor of $10^{4}$ the temperature in the centre of the Sun, $\sim 10^{7} K$. On the other hand, for the so-called \textit{stellar black holes}, whose masses are of the order of the solar mass, $M_{BH} \sim 10^{30} kg$, it is found that $T_{H} \sim 10^{-7} K$, a temperature very close to absolute zero and undetectable by astronomical observations.\\

The least massive black holes for which there is astronomical evidence are the stellar holes. The other class of black holes that have been observed are the supermassive ones, which inhabit the centre of most galaxies. These colossal objects exceed the mass of a star hole by at least a factor of $\sim 10^{6}$, which means that their temperature will be at least a factor $\sim 10^{6}$ smaller. Therefore, in stellar and supermassive holes, the quantum effects predicted by Hawking are virtually non-existent [3].

\section{Further comments: Quantum gravity and Hawking temperature}
The equation for $T_{H}$ represented the first successful step towards the elaboration of a consistent theory capable of reconciling general relativity and quantum mechanics. This theory is currently in the early stages of its development, and is known as \textit{quantum gravity}. The detailed formulation of this theory is one of the great challenges of contemporary theoretical physics.\\

To take the first step towards quantum gravity, Hawking used an approach called \textit{quantum field theory in curved spacetime}. A key aspect of this approach is that it combines general relativity (a theory of gravity) with quantum field theory\footnote{This formalism is a combination of the special theory of relativity with standard quantum mechanics.} (a formalism applicable when the effects of gravity are very small or zero). This approach has important similarities with the procedure used in the heuristic derivation of $T_{H}$, where we introduced the uncertainty principle from quantum mechanics, that is strictly only  valid in the absence of gravity, and combine it with the concept of the Schwarzschild radius from general relativity. Although Hawking's approach may seem inconsistent, this way of proceeding is common in physics, where phenomena often appear which description requires theories that are nonexistent or are in the process of being elaborated. When this happens, it is necessary to intelligently combine the different available theories, which generally correspond to seemingly incompatible physical models; this approach gives rise to results that are approximate and of limited validity, but which allow physics to move forward.\\

This is precisely the situation with the equation for $T_{H}$. Recall that this equation is valid as long as the mass of the hole is not too small, since when $M_{BH}$ approaches zero, $T_{H}$ tends to infinity. The estimated minimum mass below which the Hawking equation would no longer be adecuate is the P\textit{lanck mass}, $m_{P}$, which is defined by combining the fundamental physical constants of general relativity ($c$ and $G$) and quantum mechanics ($\hbar$):

\begin{equation}
m_{P} \equiv \left(\dfrac{\hbar c}{G}\right)^{1/2} \sim 10^{-8} kg
\end{equation}

According to the modern interpretation, $m_{P}$ is the maximum amount of mass that can be placed in a region of space with a radius of the order of the \textit{Planck length}, $l_{P}$, which is defined by combining the same constants that define the Planck mass [14]: 

\begin{equation}
l_{P} \equiv \left(\dfrac{G \hbar}{c^{3}}\right)^{1/2} \sim 10^{-35} m
\end{equation}

This is considered to be the smallest unit of distance to which a physical meaning can be attributed [14]. To appreciate how small $l_{P}$ is, remember that the typical radius of an atomic nucleus is $10^{-15} m$, which is a factor $\sim 10^{20}$ greater than $l_{P}$. The fact that $m_{P}$ and $l_{P}$ are defined by combining the characteristic constants of quantum mechanics and general relativity reveals a close connection with quantum gravity. In fact, $m_{P}$, $l_{P}$ and other analogous units constructed by combining $\hbar$, $c$ and $G$ form the so-called \textit{Planck scale}, which defines the basic units of measurement for a theory that seeks to reconcile general relativity and quantum mechanics. The Planck scale therefore determines the limits of validity of the Hawking equation, and defines the realm of reality whose description requires a detailed theory of quantum gravity.\\

In the unexplored domain of quantum gravity, quantum holes play a fundamental role, because these are small and heavy objects whose description requires reconciling quantum mechanics with general relativity. A simple calculation reveals the deep connection between black holes and quantum gravity. If we take $M_{BH} = m_{P}$ in Equation (1), we obtain $R_{S} = 2l_{P} \approx l_{P}$. Although we are wading into deep waters, this result suggests that an object of mass $m_{P}$ and radius $l_{P}$ is the smallest black hole that can be formed. It is not difficult to show that other Planck units admit a similar physical interpretation, since they define quantities that characterise the smallest quantum hole. For example, the \textit{Planck temperature} is defined as:

\begin{equation}
T_{P} \equiv \dfrac{m_{P} c^{2}}{k} = \left(\dfrac{\hbar c^{5}} {Gk^{2}} \right)^{1/2} \sim 10^{32} K
\end{equation}

This quantity can be interpreted as the Hawking temperature of the smallest black hole [14], which explains its high value.\\

All of these ideas are part of a new and fascinating field of research that is in a state of rapid development, and which took its first steps when Hawking suggested that black holes have temperature and emit radiation. However, experimental advances in this field have been slow, and for the moment it does not seem possible that great progress can be made without guidance from empirical evidence. In any case, it is not risky to say that with the passage of time, and to the extent that the physical implications are fully understood, Hawking’s finding will be considered one of the great scientific revolutions of the 20th century.

\section*{Acknowledgments}
I would like to thank to Daniela Balieiro and Michael Van Sint Jan for their valuable comments in the writing of this paper. 

\section*{References}

[1] A. Einstein, Die Grundlage der allgemeinen Relativitätstheorie, Annalen der Physik, 354 (1916) 769-822.

\vspace{2mm}

[2] A. Einstein, The Collected Papers of Albert Einstein, Princeton University Press, Princeton, 1997.

\vspace{2mm}

[3] V.P. Frolov, A. Zelnikov, Introduction to Black Hole Physics, Oxford University Press, Oxford, 2011.

\vspace{2mm}

[4] S.W. Hawking, A brief history of time, Bantam Books, New York, 1998.

\vspace{2mm}

[5] S.W. Hawking, Black Hole explosions?, Nature, 248 (1974) 30-31.

\vspace{2mm}

[6] S.W. Hawking, Particle creation by black holes, Communications in Mathematical Physics, 43 (1975) 199-220.

\vspace{2mm}

[7] N.D. Birrel, P.C.W. Davies, Quantum Fields in Curved Space, Cambridge University Press, Cambridge, 1982.

\vspace{2mm}

[8] S.B. Giddings, Hawking radiation, the Stefan–Boltzmann law, and unitarization, Physics Letters B, 754 (2016) 39-42.

\vspace{2mm}

[9] V.P. Frolov, I.D. Novikov, Black Hole Physics: Basic Concepts and New Developments, Springer Science, Denver, 1998.

\vspace{2mm}

[10] R. J. Adler, The Generalized Uncertainty Principle and Black Hole Remnants, General Relativity and Gravitation 33 (2001) 2101-2108.

\vspace{2mm}

[11] K. Krane, Modern Physics, 3 ed., John Wiley and Sons, Hoboken, 2012.

\vspace{2mm}

[12] P.A. Tipler, R.A. Llewellyn, Modern Physics, 6 ed., W. H. Freeman and Company, New York, 2012.

\vspace{2mm}

[13] M.K. Parikh, F. Wilczek, Hawking Radiation As Tunneling, Physical Review Letters, 85 (2000) 5042-5045.

\vspace{2mm}

[14] R.J. Adler, Six easy roads to the Planck scale, American Journal of Physics, 78 (2010) 925-932.

\end{document}